\documentclass[12pt]{article} 

\usepackage{graphicx,epsf} 

\def\ga{\mathrel{\raise.3ex\hbox{$>$\kern-.75em\lower1ex\hbox{$\sim$}}}}
\def\la{\mathrel{\raise.3ex\hbox{$<$\kern-.75em\lower1ex\hbox{$\sim$}}}}
\def\beq{\begin{equation}} 
\def\eeq{\end{equation}} 
\def\C{{\cal C}} 
 
\def\p{{\bf p}} 
\def\fem{f_{(em)}} 
\def\E{{\cal E}} 
\def\P{{\cal P}} 
\def\tp{{\tilde p}} 
\def\tH{{\tilde H}} 
\def\trho{{\tilde \rho}} 
\def\V{{\cal V}} 
\def\T{{\cal T}} 
\def\sigmaB{{\sigma_B}}

\title{Bulk gravitons from a cosmological brane}
\author{David Langlois\\
{\small {\it GReCO, Institut d'Astrophysique de Paris, CNRS,}}\\
{\small {\it 98bis Boulevard Arago, 75014 Paris, France}}\\
{\small and}\\
Lorenzo Sorbo\\
{\small {\it Laboratoire de Physique Th\'eorique LAPTH,}}\\
{\small {\it 9, Chemin de Bellevue, B.P. 110, 74941 Annecy-le-Vieux, France}}}

\date{\today}

\begin{document}

\maketitle 

\begin{abstract} 
We investigate the emission of gravitons  by a cosmological  
brane into an Anti de Sitter five-dimensional bulk spacetime.  We 
focus on the distribution of gravitons in the bulk 
and the associated production of `dark radiation' in this process.
In order to evaluate precisely the amount of dark radiation in the 
late low-energy regime, corresponding to standard cosmology, 
we study numerically  the emission, propagation and 
bouncing off the brane  of bulk gravitons.
\end{abstract} 

\begin{flushright}
{\small {Preprint LAPTH-989/03}}
\end{flushright}
                            
\section{Introduction} 

In the last few years, a new cosmological scenario
 has emerged, based on the  assumption that our universe is  
a {\it brane}: a sub-space embedded in a larger, bulk space, 
with additional dimensions. 
In this context, a model that has attracted particular  
attention is that of a self-gravitating three-brane embedded in an empty  
five-dimensional spacetime \cite{bdl99}, and especially its extension  
\cite{cosmors,bdel99} 
inspired by the (non cosmological) 
 Randall-Sundrum model \cite{rs99b}. 
 
In the latter case, where the bulk is endowed with  a negative  
cosmological constant, it is possible to  find
 a viable cosmological model, compatible  
so far with the available data. Due to the cosmological symmetries, it 
can be shown that in this model 
the most general bulk geometry corresponds to a portion of five-dimensional  
Anti de Sitter-Schwarzschild (AdS-Sch) spacetime, described by the metric 
\beq 
ds^2=-\left(k+\mu^2r^2-{\C\over r^2}\right)dt^2 
+\left(k+\mu^2r^2-{\C\over r^2}\right)^{-1} dr^2+r^2d\Sigma_k^2, 
\label{sch-ads} 
\eeq 
with $k=-1,\,0,\,1$ depending on the curvature of 
the three-dimensional spatial  
slices. The  bulk cosmological constant is related to the mass scale 
$\mu$ via  
\beq 
\Lambda=-6\,\mu^2. 
\eeq

From the point of view of the brane, whose energy 
density is supposed to be the sum of  
an intrinsic tension $\sigma$ and of the usual 
cosmological matter energy density $\rho$, 
cosmology is governed by the brane Friedmann equation, which is  
different from the standard one and reads \cite{bdel99}
\beq 
H^2\equiv \left({\dot a\over a}\right)^2 =\left({\kappa^4\over 36}\sigma^2-\mu^2\right) 
+{\kappa^4\over 18}\sigma\rho 
+{\kappa^4\over 36}\rho^2-{k\over a^2}+{\C\over a^4}. 
\label{bdel} 
\eeq 
This result can be obtained by applying the gravitational junction conditions 
for a moving  brane  
in (\ref{sch-ads}), the cosmological scale factor $a(t)$ being 
simply the radial coordinate $r$ at the brane position.
The last term in eq.~(\ref{bdel}), $\C/a^4$, is usually called 
``dark radiation" 
or ``Weyl radiation" and represents the influence of the bulk 
geometry on the 
cosmology in the brane. 
 
If the bulk is strictly AdS-Sch then $\C$ is  a constant and corresponds to
 the five-dimensional  
mass in the Schwarzschild AdS metric (\ref{sch-ads}).  
 The   term $\C/a^4$
then behaves exactly as a radiation component. 
The main constraint on the amount of 
such extra radiation comes from nucleosynthesis,  
since the primordial abundances of light elements depend crucially on the  
balance between the expansion  rate of the universe 
and the rate of  microphysical  
reactions.  
  The good agreement of observations 
with the predictions of standard nucleosynthesis implies an 
upper bound on the number of nonstandard relativistic degrees of 
freedom in the universe and thus, in the context of brane 
cosmology, an upper bound on the value of the Weyl parameter $\C$.

Now, if the bulk is not strictly empty but contains a non-vanishing bulk  
energy-momentum tensor, then the Weyl parameter $\C$ is no longer necessarily 
constant. In other words, the generalization of 
Birkhoff's theorem to brane cosmology no longer applies. 
This has been  
illustrated in particular  in models with a bulk 
scalar field \cite{lr01,charmousis01}. 
 
In fact, in a realistic brane cosmological scenario,  
there is an unavoidable bulk component, which is simply due to the  
 the gravitational waves, or bulk gravitons, produced  
by the brane matter fluctuations. Such 
 bulk gravitons are created by the scattering of ordinary  
matter particles confined on the brane.  
 The expected consequence  of this phenomenon is to `feed' the Weyl parameter
 by means of a transfer of energy 
from the brane into the bulk. 

This problem has been recently investigated in \cite{hm,lsr,kkttz}.
 Whereas the authors 
of \cite{kkttz}
 analyze generic forms of transfer of energy between brane and bulk 
using a four dimensional effective description, the case of pure graviton 
emission has been studied in HM \cite{hm} and LSR \cite{lsr}.

In HM, the authors consider a pure AdS bulk and analyze the 
generation of dark radiation by considering two distinct phases. In the 
low energy regime ($\rho\ll \sigma$) the amount of generated dark 
radiation is obtained by directly  equating it to the loss of 
energy density on the brane. In the high energy regime, the 
emitted gravitons are considered to  remain
gravitationally bound to the brane for the whole duration of this era, 
bouncing several times off it. 
To estimate the 
corresponding amount of dark radiation, the lost energy is reduced  
by a  factor due  to the energy dissipation resulting  
 from the repetitive  collisions with the brane.
Because of the somehow crude
 distinction between the 
high and low energy eras, this analysis cannot follow the details 
of the transition between these two regimes. Moreover the authors of HM
give their final result up to an uncertainty factor, estimated to be 
between $0.5$ and $1$, due to the multiple bouncing of gravitons off the brane.

In LSR, 
a five--dimensional {\it exact} solution for the bulk is used, taking 
into account     an energy flux from the brane into the bulk. 
The simplest solution of this type 
is the five-dimensional generalization of Vaidya's metric. 
In this way the brane trajectory, and thus its cosmological evolution, 
are determined self--consistently via  the junction conditions, 
taking into account 
possible backreaction effects. The price to be paid for this is the necessity 
to make the assumption that the trajectories followed by the bulk 
gravitons are exactly perpendicular to the brane, assumption that 
is realistic only in the low energy regime.

Although these two approaches are very different, they give estimates  
that agree within one order of magnitude. Moreover, these estimates 
are very close to  the current observational bound on the number 
of extra relativistic degrees of freedom. 
Therefore, it is important to get an accurate 
determination of the amount of dark radiation expected for this model. Indeed, 
the improvement of observational constraints on the number of non standard 
relativistic degrees of freedom will allow us to put constraints on
the cosmological evolution of the Randall-Sundrum model (possibly excluding 
a long era of nonstandard $H^2\propto \rho^2$ cosmology) and on the 
parameter space of the model.
A detailed analysis can also allow us to quantify the effects of the 
bounces of gravitons on the brane and to estimate to which extent 
backreaction effects can be relevant. The present analysis, finally, can be 
generalized to the case of models with extra matter in 
the bulk \cite{kkttz,ht03}, eventually leading to stronger constraints.

Our plan is the following. In the next section, we discuss the problem
from an effective four-dimensional point of view. In section 3, we rederive
the equations governing the cosmology of the brane and the trajectory 
of the brane in the bulk. Section 4 is devoted to the emission of bulk 
gravitons by brane particles. In section 5, we recall our model  \cite{lsr}
based on the Vaidya metric. In section 6, we discuss the propagation of 
the gravitons in the bulk. In section 7, we present and 
discuss our numerical computations. We conclude in the final section.

\section{Effective approach} 
As a starting point, we will discuss the cosmology of the brane and  
the influence of the bulk gravitons from a four-dimensional effective  
description. 
In order to do so, we will follow 
the approach of \cite{sms99}.

We start with the 5-dimensional Einstein equations 
\begin{equation} 
  R_{AB}-{1\over2}g_{AB}R+\Lambda_5g_{AB} 
=\kappa^2\left[{\T}_{AB}+S_{AB} \delta(y)\right]\,, 
\label{einstein} 
\end{equation} 
where the matter component, on the right hand side, consists of a  
bulk energy-momentum tensor $\T_{AB}$ and 
distributional a brane energy-momentum tensor. The brane 
is located at  $y=0$, $y$ corresponding to  
the proper coordinate normal to the brane. We assume that
the extra dimension has a $Z_2$ orbifold symmetry and that $y=0$ is 
a fixed point under this  symmetry.

In the cosmological extension of  
the Randall-Sundrum  model \cite{rs99b}, 
$S_{AB}$ is the sum of a  brane tension $\sigma$, defined by 
\beq 
\kappa^2\sigma=\sqrt{-6\Lambda}=6\mu, 
\label{tension} 
\eeq 
and of 
the contribution of ordinary matter fields  
confined to the brane $\tau_{AB}$, i.e. 
\begin{equation} 
S_{AB} = -\sigma h_{AB}+\tau_{AB}\,, 
\label{matter} 
\end{equation} 
where $h_{AB}$ is the metric induced on the brane. 
 
From the above  
five-dimensional Einstein's equation, using the $Z_2$ symmetry,  
one can derive \cite{sms99} 
effective four-dimensional equations that read 
\begin{equation} 
  {}^{(4)}G_{\mu\nu}= 
\kappa_4^2(\tau_{\mu\nu} 
           +\tau_{\mu\nu}^{(\pi)}+\tau_{\mu\nu}^{(W)} 
+\tau^{(B)}_{\mu\nu} 
            ),   
\label{4dEinstein} 
\end{equation} 
with $\kappa_{4}^{2} = {\kappa^2\mu}$  
and  
\begin{eqnarray} 
\kappa_{4}^{2}\tau_{\mu \nu}^{(\pi)} &=&  
 -{\kappa^2 \over 24}\left[6\tau_{\mu \alpha}\tau_{\nu}^{\ \alpha} 
 -2\tau \tau_{\mu \nu}  
 -h_{\mu \nu}(3\tau_{\alpha \beta}\tau^{\alpha\beta} 
  -\tau^{2})\right],\hspace{-1cm} 
\nonumber\\ 
\kappa_{4}^{2}\tau_{\mu \nu}^{(W)}&=& \,- 
  {}^{(5)}C_{ABCD} 
\,n^{A}\,n^{B}\,h_{\mu}^{B}\,h_{\nu}^{D}\,, 
\nonumber\\ 
 \kappa_{4}^{2} \tau_{\mu\nu}^{(B)} &= &{2\kappa^2\over 3} 
   \left[{\T}_{AB} h^A{}_\mu h^B{}_\nu 
          +h_{\mu\nu}\left({\T}_{AB} n^A n^B 
            -{1\over 4} {\T}^{A}{}_A 
             \right)\right],\hspace{-0cm} 
\label{eq4} 
\end{eqnarray} 
where $n^{A}$ is the unit vector pointing outward and   
normal to the brane. In addition to the usual  
four-dimensional matter energy-momentum tensor $\tau_{\mu\nu}$,  
three new terms appear on the right hand side of the effective  
Einstein equations: the first one 
is quadratic in $\tau_{\mu\nu}$; the second one is the projection on the brane 
of the 5-dimensional Weyl tensor 
${}^{(5)}C_{ABCD}$; and  the last one is the  
projected effect of the bulk energy-momentum tensor. 
 
Since we are interested here in homogeneous brane cosmology, all the  
effective energy-momenta defined above have necessarily  
a perfect fluid structure, i.e. 
\beq 
 \tau^{(i)}_{\mu\nu}=\left(\rho^{(i)}+p^{(i)}\right)u_\mu u_\nu 
+p^{(i)}h_{\mu\nu}, 
\eeq 
where $u^\mu$ is the timelike unit vector associated with comoving 
observers on the brane. 
 
It is then not difficult to show that  Eq.~(\ref{4dEinstein}) 
gives a Friedmann equation on the brane of the form  
\begin{eqnarray} 
H^2 =   
{{\kappa _4^2} \over 3}\left[  
\left(1+{\rho\over 2\sigma}\right)\rho+\rho^{(W)}  
 +\rho^{(B)}\right],  
\label{Hubble} 
\end{eqnarray} 
where $H=\dot a/a$ is the Hubble parameter on the brane ($a$ is the scale  
factor on the brane and the overdot stands for a  derivative  
with respect to the cosmic proper time $t$).  
We have used the relation $\rho^{(\pi)}= \rho^2/2\sigma$.  
 
Let us now discuss the various (non) conservation laws satisfied 
by the different energy density components defined above.
We follow here the recent analysis of \cite{ht03}, where the bulk 
energy-momentum tensor was associated to a five-dimensional scalar field.
One can first show that  
the usual conservation equation for cosmological matter is modified  
into   
\begin{equation}\label{flux} 
\dot{\rho}+3\,H\,\left(\rho+p\right)=2\,{\cal {T}}_{RS}\,n^R\,u^S\,\,. 
\end{equation} 
The right-hand side (evaluated at the brane position)  
represents the energy flux {\it from} the bulk {\it into} 
the brane. When the brane {\it loses} energy, as will be the case here via 
emission of gravitons,  the right-hand side is {\it negative}. 
The factor $2$ is a consequence of the $Z_2$ symmetry. 
 
Moreover, the 4-dimensional Bianchi identities  
imply that the total energy density defined by  
$\rho^{({\rm{tot}})}=\rho+\rho^{(\pi)}+\rho^{(B)}+\rho^{(E)}$  
satisfies 
\begin{equation} 
 \dot\rho^{({\rm{tot}})}+ 4H\rho^{({\rm{tot}})} 
+H \tau^{({\rm{tot}})\mu}{}_{\mu}=0. 
\end{equation}  
Using  
$\tau^{(\pi)\mu}{}_{\mu}= {(\rho/\sigma)}(\tau^{\mu}{}_{\mu}+2\rho)$,  
$ \tau^{(B)\mu}{}_{\mu} = 2 (\kappa^2/\kappa_{4}^{2}) {\T}_{AB} n^A n^B$,
$ \tau^{(W)\mu}{}_{\mu} = 0$ and the (non) conservation equation~(\ref{flux}), it follows 
that the energy density for the {\it dark component}, by which we mean  
the sum of the components depending explicitly on the bulk,  
i.e. 
$\rho_{\rm{D}}=\rho^{(B)}+\rho^{(W)}$, satisfies   
\begin{equation} 
 \dot \rho_{\rm{D}}+4H\rho_{\rm{D}} 
  =-2\left(1+{\rho\over\sigma}\right){\T}_{AB} u^A n^B   
  -2 {H\over \mu} \, {\T}_{AB} n^A n^B\,, 
\label{evol_DR} 
\end{equation} 
where the right hand side is evaluated at the brane position. 
On the right-hand side of the  
above equation, one recognizes in the first term   
 the energy flux {\it from} the brane {\it into} the bulk,  
$-2\,{\T}_{AB} u^A n^B$. This means, not surprisingly, that  
 the loss of energy inside the brane, will contribute to an  
{\it increase} of the amount of dark radiation. In the second  
term, the quantity  
${\T}_{AB} n^A n^B$ can be interpreted  
as the  {\it pressure} transverse to the brane, due to the bulk component.  
In the case of a gas of gravitons, which we consider here, this  
pressure is positive and therefore 
this term tends to {\it decrease} the amount of dark radiation. The two 
terms on the right hand side have thus opposite effects.

\section{Cosmology of a brane  in AdS} 
We now consider explicitly the bulk and  
 study the trajectory of a   brane with relativistic matter 
in a strictly AdS  spacetime,  with the  metric 
\beq 
ds^2=-f\left(r\right)\,dT^2+\frac{dr^2}{f\left(r\right)}+r^2\,d{\bf x}^2,\quad 
f(r)=\mu^2\,r^2. 
\label{metric} 
\eeq 
The trajectory of the brane can be represented in terms of its coordinates  
$T(t)$ and $r(t)$ as functions of the proper time $t$, which is also the  
cosmic time in the brane. 
The normalization of the velocity vector $u^A=(\dot T, \dot r, {\bf 0})$  
then implies 
\beq  
u^A=\left(\frac{\sqrt{f+\dot{r}^2}}{f},\,\dot{r},\,{\bf 0}\right)\,\,. 
\eeq 
One then needs the junction conditions for the brane, 
\beq 
\left[h_A^C\nabla_C n_B\right]=\kappa^2\left(\tau_{AB}-{1\over 3}\tau  
h_{AB}\right),  
\eeq 
where $n^A$ is  the unit vector  
normal to the brane (pointing outwards) and is given by 
\beq 
n^A=-\left(\frac{\dot{r}}{f},\,\sqrt{f+\dot{r}^2},\,{\bf 0}\right)\,\,. 
\eeq 
The spatial components of the junction equations yield  
 the brane Friedmann equation, which can be expressed as  
\beq 
\frac{H^2}{\mu^2}=2\,\frac{\rho}{\sigma}+\frac{\rho^2}{\sigma^2}. 
\label{fried}
\eeq 
 
It is convenient to introduce the dimensionless quantities  
\beq 
\tH={H\over \mu}, \qquad \trho={\rho\over\sigma}. 
\label{tilde} 
\eeq 
Note, using (\ref{tension}) and $\kappa_4^2=\kappa^2\mu$  
 that the  brane tension can be expressed as  
\beq 
\sigma=6\mu^2 m_P^2, 
\eeq 
where $m_P=1/\sqrt{\kappa_4}$ is the reduced Planck mass. 
We will drop the tildes from now on. 
 
In terms of $H$, the components of the  brane velocity and of the  
normal vector  read respectively  
\beq 
u^A=\left({1\over r}\sqrt{1+H^2}, rH, {\bf 0}\right), 
\qquad 
n^A=-\left({H\over r}, r\sqrt{1+H^2}\right). 
\label{un} 
\eeq 
The trajectory of the brane is obtained by solving  
\beq 
{dr\over dT}={\dot r\over \dot T}=r^2{H\over \sqrt{1+H^2}}= 
r^2{\sqrt{2\rho+\rho^2}\over 1+\rho},
\eeq 
where the second equality follows from (\ref{un}) and the third from 
(\ref{fried}).
In the case of radiation domination, $\rho=\rho_i\,r^{-4}$, where we have 
defined the radial coordinate $r$ such that $r=1$ for $\rho=\rho_i$, where 
$\rho_i$ is the energy density at some fiducial initial time $t_i$. 
To get the brane trajectory, one must integrate 
\beq\label{branetra} 
{dr\over dT}=r^2{\sqrt{2\,\rho_i\,r^4+\rho_i^2}\over r^4+\rho_i}. 
\eeq 
Explicit integration is possible 
in the high energy regime, $\rho_i/r^4\gg 1$,  
where one finds 
\beq  
{1\over r}\simeq -T +{\mathrm {const.}} 
\eeq 
and in the low energy regime, where
\beq 
r\simeq \sqrt{2\,\rho_i}\, T + {\mathrm {const.}} 
\eeq 
In the regime of transition between the high energy phase and the low energy  
phase, i.e. for $r\sim \rho_i^{-1/4}$, one must resort to numerical integration.

\section{Emission of bulk gravitons} 
We now consider the production of bulk gravitons by the cosmologically  
evolving brane. In particular, we wish   
  to compute explicitly the components 
of the bulk energy-momentum tensor due to the gas of gravitons 
emitted by the brane. 
 
Let us first recall that the energy-momentum tensor due to a gas of  
massless particles is given by
\beq 
\T_{AB}=\int d^5p \ \delta\left(p_Mp^M\right)\sqrt{-g}\,f\, p_Ap_B, 
\label{T_integ} 
\eeq  
where $f=f(x^A,p_A)$ is the distribution function of the gravitons
(which is a scalar that depends on the spacetime position and on the 
momentum).
 
Since the gravitons living in the bulk will be assumed to have been produced  
only by emission from the brane, it will be convenient to start the  
computation of the energy-momentum tensor components on the trajectory of the 
brane. At the location of the brane, we can see the bulk gravitons  
from two perspectives.  
 
First, from the brane point of view, 
the bulk gravitons are seen as massive four-dimensional particles  
with  mass $m$,  three-momentum $\p$ and  energy $E$, satisfying   
$E=\sqrt{\p^2+m^2}$. 
They  are created  
by the scattering of two ordinary particles confined on the brane.  
The leading contribution to this process is
 given by the scattering $\psi\,\bar{\psi}  
\rightarrow$ graviton, where $\psi$ is a standard model particle. 
 At the cosmological level, the production of  
gravitons results into an energy loss for ordinary matter, which can be 
 expressed as  
\beq 
{d\rho \over dt}+ 3H(\rho+P)=- 
\int\frac{d^3p}{\left(2 \pi\right)^3} 
{\bf C}\left[f\right],  
\label{prl16} 
\eeq 
with  
\beq 
{\bf C}\left[f\right]={1\over 2} 
\int \frac{d^3p_1}{\left(2 \pi\right)^3\,2E_1}\,  
\frac{d^3p_2}{\left(2 \pi\right)^3\,2E_2} 
\,\sum \left\vert {\cal {M}}\right\vert^2\,f_1\,f_2\, 
\left(2\pi\right)^4\,\delta^{(4)}\left(p_1+p_2-p\right)\,\, , 
\label{prl17} 
\eeq 
where ${\cal {M}}$ is the scattering amplitude for the process in  
consideration, and the indices $1$ and $2$ correspond to the scattering 
particles ($\psi$ and $\bar\psi$).  
 
Second, from the bulk point of view, the gravitons are  
massless particles, each graviton being characterized by a five-dimensional  
momentum, which can be decomposed into  a spatial four-momentum and an energy, 
 defined with respect to a reference frame.  
 
To make the connection between these two points of view, it is convenient to  
choose a reference frame associated to a (comoving) brane observer. 
Introducing an orthonormal frame defined by $u^A$, $n^A$ and $e_i^A$,  
the five-dimensional momentum of any graviton can be decomposed into  
\beq 
p^A= E u^A+mn^A+\sum_i \tp_i e_i^A. 
\label{decomp_brane} 
\eeq 
The components along the vectors $e_i^A$ are denoted with a tilde to  
distinguish them from the components $p^i$ defined in the decomposition 
along the coordinate vectors $(\partial/\partial x^i)^A$. 
If one substitutes this decomposition into (\ref{T_integ}), 
 one gets for the mixed component  
\beq 
\T_{nu}\equiv u^An^B\T_{AB}=\int d^5p\, \delta\left(p_M\,p^M\right) 
\,\sqrt{-g}\,f\,p_Ap_B\,n^Au^B=-\int dm\, d^3{\bf{p}}\,\frac{m}{2}\,f\,\,. 
\eeq 
By identifying the right hand sides of (\ref{flux}) and (\ref{prl16}),  
one immediately finds, upon  
comparing (\ref{prl17}) with the above expression,  
 that the distribution function at the brane location,  
for gravitons which are being emitted,  is given by  
\beq 
\fem\left(m,\,{\bf p}\right)=\frac{1}{2\,m}\,\frac{1}{\left(2\,\pi\right)^5} 
\,\int\frac{d^3{\bf p}_1}{2\, p_1}\,\frac{d^3{\bf p}_2}{2\, p_2}\, 
\sum\left| \cal{M}\right|^2\,f_1\,f_2\,\delta^{(4)}\left(p_1+p_2-p\right). 
\eeq 
The summed squared amplitude is given by the expression    
\begin{equation} 
\sum \left\vert \cal{M}\right\vert^2=A\,\frac{\kappa^2}{8\,\pi}\,s^2\,\,  
\end{equation} 
where  $s$ is the Mandelstam invariant and  
\beq 
A={2\over 3}g_s +g_f+4g_v, 
\eeq 
where $g_s$, $g_f$ and $g_v$ are respectively the scalar, fermion and  
vector relativistic degrees of freedom in thermal equilibrium on the brane. 
For the standard model, if all degrees of freedom are relativistic, one 
has $g_s=4$, $g_f=90$ and $g_v=24$.  
For simplicity, we will ignore the Bose  
or Fermi corrections in the distribution functions and assume that  the  
particles on the brane are characterized by  a Boltzmann distribution 
of temperature $T$. The distribution function at emission, $\fem$, can then 
 be  
computed explicitly and  
one finds (neglecting the masses of the scattering particles) 
\begin{equation}\label{f_em} 
\fem\left(m,\,{\bf p}\right)= 
\frac{A}{2^{10}\,\pi^5}\,\kappa^2\,m^3\,e^{-\sqrt{{\bf p}^2+m^2}/T}\,\,. 
\end{equation} 
 
We now have all the elements to compute all the components of the bulk  
energy-momentum tensor due to the emitted gravitons at the brane location. 
They are given by 
\begin{eqnarray}
\T^{(em)}_{uu}&=& \int dm\, d^3{\bf{p}}\,\frac{E}{2}\,\fem= 
{21\over 16 \pi^4}A\kappa^2 T^8= 
{4725\over 4\pi^8}{A\over g_*^2}\kappa^2\rho^2, \label{Tuu}
\\ 
\T^{(em)}_{un}&=& -\int dm\, d^3{\bf{p}}\,\frac{m}{2}\,\fem= 
-\frac{315\, A}{2^{10}\,\pi^3}\,\kappa^2\,T^8 
=-{70875\over 2^8\pi^7}{A\over g_*^2}\kappa^2\rho^2, 
\label{Tnu}
\\ 
\T^{(em)}_{nn}&=& \int dm\, d^3{\bf{p}}\,\frac{m^2}{2E}\,\fem= 
\frac{3\, A}{4\,\pi^4}\,\kappa^2\,T^8 
={675\over \pi^8}{A\over g_*^2}\kappa^2\rho^2, 
\label{Tnn}
\end{eqnarray} 
with $E\equiv \sqrt{\p^2+m^2}$ as stated before. In the second equalities,  
we have replaced the temperature by the energy density, using 
$\rho=(\pi^2/30)g_*T^4$, where $g_*$ is the effective number of  
relativistic degrees of freedom.  
The above components can be interpreted physically respectively as the  
energy density, the energy flux and the lateral pressure of the emitted  
bulk gravitons as measured by an observer at rest with respect to the brane.

\section{Cosmological evolution in the Vaidya model} 
In this section, we deviate from the preceding analysis and consider 
the evolution of the dark radiation in the context of the Vaidya model  
introduced in \cite{lsr} (see also \cite{ckn00}). 
The interest of the Vaidya model  
is to work with an explicit five-dimensional realization of the effective  
equations introduced in section 2, where there is a non zero energy  
transfer between the brane and the bulk. 
This however imposes 
 that all the gravitons have to be assumed to be emitted  
{\it radially} in the five-dimensional bulk so  that the  
bulk energy-momentum tensor is of the form 
\beq 
{\T}_{AB}=\sigmaB k_A k_B,  
\label{Tnull} 
\eeq 
where $k^A$ is a null vector. Assuming spherical symmetry,  
Einstein's equations (\ref{einstein}) with such an energy-momentum 
can be solved analytically: this is the Vaidya solution \cite{vaidya},  
ordinarily 
used to describe a radiating relativistic star. Its metric is given by 
\beq 
ds^2=-f\left(r,\,v\right)\,dv^2+2\,dr\,dv+r^2\,d{\vec x}^2, 
\eeq 
with  
\beq 
f(r,v)=\mu^2 r^2-{\C(v)\over r^2}. 
\eeq 
For a constant $\C$, one recovers, after changing the light-like  
coordinate $v$ into the static time coordinate $t$,  
the familiar five-dimensional AdS-Sch metric.  
 
It is instructive to apply the  effective equations derived earlier  
 in this particular context. First, one can normalize the null vector $k^A$  
such that $k_Au^A=1$, in which case the projections of the   
bulk energy-momentum tensor (\ref{Tnull}) are given by  
${\cal {T}}_{nu}=-\sigmaB$ and ${\cal {T}}_{nn}=\sigmaB$.  
As a consequence the non-conservation equation (\ref{flux})  
for the energy density on the  
brane reads 
\beq\label{enoncon} 
\dot{\rho}+3\,H\,\left(\rho+p\right)=-2\,\sigmaB, 
\eeq 
and $\sigmaB$, up to the factor of $2$ due to the $Z_2$ symmetry, directly  
represents the energy loss in the brane due to the production of bulk 
gravitons. The dependence of $\sigmaB$ is given explicitly in  
(\ref{Tnu}), and in terms of the brane energy density, 
\beq 
\sigmaB\propto \rho^2. 
\eeq

The { Weyl parameter} $\C$ can be related to the ``dark'' component  
energy density defined earlier via 
\beq 
\rho_{D}={\C\over a^4}. 
\eeq 
Substituting in  
the evolution equation (\ref{evol_DR}), one gets 
 (still using the implicit tilded quantities defined in (\ref{tilde})) 
\beq 
{\dot\C\over a^4}= 
-2\left(1+{\rho}\right){\T}_{AB} u^A n^B   
  -2 H \, {\T}_{AB} n^A n^B\,. 
\label{evolC} 
\eeq 
With the explicit bulk energy-momentum tensor (\ref{Tnull}),  
$T_{nu}=-\sigmaB$ and $T_{nn}=\sigmaB$, Eq. (\ref{evolC}) reduces to  
\beq\label{eqlsr}
{\dot\C\over a^4}= 
2\sigmaB\left(1+\rho  
  -  H\right)\,, 
\eeq 
where one recognizes  the result already derived in \cite{lsr}, but there   
  from Einstein's equations rather than from the effective equations like 
here. 
The Friedmann equation, when one neglects the dark radiation contribution,  
gives during the high energy regime 
\beq 
H\simeq \rho+1 -{1\over 4\rho}. 
\eeq 
This shows that there is a remarkably precise  
compensation, {\it at leading order and  
next to leading order}, between the energy flux and the transverse pressure 
in the dark radiation equation. And the production of $\C$ is governed  
by  
\beq 
{\dot\C\over a^4}\simeq
{\sigmaB\over 2\rho}\propto \rho. 
\eeq 
 
As mentioned above, the drawback of the Vaidya description  
is the   assumption that all gravitons are radial. 
As seen in the previous section, the distribution of emitted gravitons 
is not radial and ${\cal {T}}_{nn}+{\cal {T}}_{un}$ is not zero.  
Therefore, if one substitutes the explicit values 
of ${\cal {T}}_{nn}$ and ${\cal {T}}_{un}$ in (\ref{evolC}) we  no longer get the very  
precise compensation observed in the Vaidya case and the global  
sign is positive (since $315/(2^{10}\pi^3)>3/(4\pi^4)$), so that the  
production of dark radiation seems to be driven in the high energy regime  
by a term on the right hand  
side proportional to $\rho^3$, rather than
proportional to $\rho$ as in the Vaidya description. Obviously, this would  
result in an enormous amount of dark radiation, far above the estimate of  
\cite{lsr}, 
and this would ruin the simplest brane cosmology scenario, since the  
estimate of \cite{lsr} was barely within the nucleosynthesis bounds. 
 
The above analysis is however incomplete. 
In the Vaidya description, the gravitons 
are emitted radially inwards and are therefore  lost for the brane once  
there are emitted. Some of the non radial gravitons, however, can come back  
onto the brane after their emission, as it will be shown in the next section, 
and thus influence once more the evolution of the Weyl parameter.   
Because of the $Z_2$ symmetry, these gravitons 
will be reflected by the brane (we ignore here the decay   
of a bulk graviton into brane particles) and will contribute  only to 
 the transverse pressure term. The effect of these  
old gravitons  will thus be to reduce the amount of dark radiation 
that would be computed naively by considering only gravitons being 
emitted. 
 
\section{Graviton trajectories in the bulk} 
After the gravitons are emitted, they  move freely in the bulk, each  
individual graviton following a null geodesic.  
The null geodesics in five-dimensional AdS were studied in \cite{cl01}. 
We summarize 
here  the main results. Using the symmetries of the metric (\ref{metric}),  
one can  identify the first integrals for the geodesic  motion  
\beq 
f(r){d T\over d\lambda}=\E, \qquad  
r^2 {d x^i\over d\lambda}=\P^i, 
\eeq 
where $\lambda$ is any affine parameter.   
For any graviton one can choose the affine parameter so that the  
tangent vector of the null trajectory is identified with the physical  
momentum $p^A$.  
Introducing the notation $\tp^T=rp^T$ and $\tp^r=p^r/r$, the above first  
integrals become 
\beq 
\E=r \tp^T 
\label{conserv_E} 
\eeq 
and  
\beq 
{\bf }\P=r\p. 
\label{conserv_P} 
\eeq 
 
Using these conservation laws, it is easy to determine the trajectory  
of the gravitons in the bulk spacetime. In order to do so, one  
can compute 
\beq 
{dr\over dT}=\left({d r\over d\lambda}\right)/\left({dT\over d\lambda}\right). 
\eeq 
Since $p^Ap_A=0$, one gets  
\beq  
\left(p^r\right)^2=\E^2-\P^2, 
\eeq 
which means that $p^r$ is also a constant of motion. 
It is then useful to introduce the parameter 
\beq 
\V={p^r\over \E}, 
\eeq 
which will be constant along the null geodesic trajectory, and in terms  
of which the trajectory in the time/extra-dimension submanifold is given 
by 
\beq  
{dr\over dT}=\V r^2. 
\eeq 
Integration gives the graviton trajectory 
\beq\label{nullgeo} 
T-T_*=-{1\over\V}\left({1\over r}-{1\over r_*}\right). 
\eeq 
 
It is also useful to relate the  bulk-based  description used above  
 with the brane-based approach introduced earlier. Using the  
decomposition  (\ref{decomp_brane}) with the  velocity and normal  
vectors given in (\ref{un}), one finds 
\beq 
\tp^T=\sqrt{1+H^2}\, E - H m, \qquad \tp^r=H \, E-\sqrt{1+H^2}\,  m. 
\eeq 
It is immediate to invert this system, to get 
\beq 
m= H \tp^T- \sqrt{1+H^2}\, \tp^r, \qquad 
E=  \sqrt{1+H^2}\,  \tp^T - H\,  \tp^r. 
\eeq 
 
For some values of the parameters, the trajectory of the graviton, leaving  
the brane at the emission point, can cross again the brane trajectory.  
It will then be reflected by the brane, as a consequence of the  
$Z_2$ symmetry.  
 
It is not difficult to compute the new five-dimensional momentum of the  
graviton after the reflection on the brane, by  
considering the reflection from the brane point of view. The momentum  
parallel to the brane $\p$ is conserved, as well as the energy $E$. Only  
the transverse momentum is affected and simply changes its sign. The 
momentum along the spatial direction orthogonal to the brane is embodied 
by the mass $m$. We will adopt the convention that $m$ is positive when  
the momentum is outwards, with respect to the brane, and negative  
otherwise.  
The reflection of the graviton by the brane is thus governed by the  
simple laws, 
\beq 
E\rightarrow E, \quad  \p \rightarrow \p, \quad m\rightarrow -m. 
\eeq 
In the bulk-based point of view, this translates into 
\begin{eqnarray} 
\tp^T_{out}&=&(1+2H^2)\, \tp^T_{in}-2H\sqrt{1+H^2}\, \tp^r_{in}, 
\\ 
\tp^r_{out}&=&2H\sqrt{1+H^2}\, \tp^T_{in}-(1+2H^2)\, \tp^r_{in}. 
\end{eqnarray}

We can use the previous results to infer the  
 evolution of the graviton distribution function throughout the bulk. 
In principle, it is governed  by the  
Liouville equation, which, in general relativity, reads  
\beq 
p^A{\partial f\over \partial x^A}+\Gamma_{AB}^Cp^Ap^B {\partial f\over  
\partial{p^C}}=0. 
\label{liouville} 
\eeq 
However, this 
equation is in fact no more than the implementation at the level  
of the distribution function of the fact that 
each individual particle follows  
a geodesic. To solve explicitly the Liouville equation it is therefore 
 simpler to use directly the solutions for the geodesic trajectories.  
 
This enables us to write the emission distribution function, at the brane  
location, in terms of the bulk-based momenta, by simply substituting the  
above expressions in (\ref{f_em}). 
Moreover, using the constants of motion along the graviton geodesics  
established above, one can deduce the expression for the distribution  
function off the brane as well. One gets 
\beq 
\fem^{bulk}(T,r,\tp^T,\tp^r,\p)= 
\fem(T_{em},a_{em}, {r\over a_{em}} \tp^T,{r\over a_{em}}\tp^r,  
{r\over a_{em}}\p), 
\eeq 
where the time and brane scale factor at emission depend on  
$t$, $r$, $\tp^T$ and $\tp^r$ and are obtained by tracing the graviton  
geodesic back in time until it crosses the brane trajectory.  
The rescaling of the momenta on the right hand side are just a consequence 
of the conservation laws (\ref{conserv_E}) and (\ref{conserv_P})  
(the same rescaling for $\tp^r$ follows from the normalization of the  
five-dimensional momentum). By construction, the above distribution function 
must be a solution of the Liouville equation (\ref{liouville}) in the bulk, 
as one can explicitly check. 
 
The above expression describes gravitons which have been emitted by the brane  
and have not returned back onto the brane. But, as already mentioned, the  
situation is in fact more complicated because some of the gravitons  
that have been emitted by the brane can come back onto the brane and 
 be reflected back into the bulk with a different momentum. Therefore,  
at each point along the brane trajectory, one must add to the gravitons  
that are being emitted all the older gravitons that are being reflected  
by the brane at the same instant.  
 
Actually, the {\it majority} of gravitons emitted in the high energy regime will
be reflected at least once by the brane. In the period in which
$H^2\propto \rho^2$ (i.e. $H \gg 1$) the brane is indeed moving
relativistically with respect to the frame defined by the metric (14). The
gravitons that are not emitted exactly orthogonal to the brane will see their
radial momentum boosted, and will thus move in this frame in the same
direction as the brane. This can be seen by inspection of eq. (56): for
$H\gg 1$ and $E$ not too close to $m$, $p^r$ is positive. As $H$
decreases, the brane will eventually slow down, and the gravitons will
bounce off it. 

Let us consider a fiducial time, characterized by $t_0$, in the history  
of the brane. In order to compute the increment of the Weyl parameter, via  
(\ref{evolC}), one must add to the newborn gravitons, which are just being  
emitted, the old gravitons that are being reflected by the brane. 
For these old gravitons, not surprisingly, the net contribution   
to the energy flux $T_{nu}$ vanishes, since they  
are just reflected, neither absorbed nor created, whereas the incoming  
and outgoing contributions to the pressure are equal in magnitude and  
opposite in sign. 
For these gravitons, therefore, we need to compute only the component  
\beq 
{\cal {T}}_{nn}^{(in)}=\int dm\, d^3{\bf{p}}\,\frac{m^2}{2E}\,f_{(in)}. 
\label{Tnn_in} 
\eeq 
 
Let us now write explicitly the contribution to $f_{in}$ from gravitons  
which make at the time $t_0$ their first bounce since their emission.  
It will be convenient to introduce the following parametrization  
\beq 
H=\sinh\beta, 
\eeq 
for the Hubble rate, and for each graviton, the parameter $x$ such that  
\beq 
E=p\cosh x, \qquad m=p\sinh x, 
\eeq 
where $p\equiv |{\bf p}|$ is  the norm of the  three-momentum parallel 
to the brane.  
Using the results obtained previously, one finds that  
the energy $E_1$ and mass $m_1$ measured in the brane frame at the  
time of emission $t_1$ are given by the simple expressions 
\beq 
E_1={a_0\over a_1}p\cosh\left(\beta_1-\beta_0+x\right), 
\qquad 
m_1={a_0\over a_1}p\sinh\left(\beta_1-\beta_0+x\right), 
\eeq 
where 
\beq 
\beta_1=\beta_1(\beta_0,x), 
\eeq 
is determined by tracing backwards the null geodesic 
(see Fig. 1)  that intersects the brane trajectory 
at $t_0$ until the previous intersection. More explicitly, $\beta_1$ 
is obtained by solving the equation 
$T_b\left(r\right)=T_g\left(r\right)$, where $T_b\left(r\right)$ gives  
the time $T$ as a function of the position $r$ of the brane through  
eq.~(\ref{branetra}), whereas $T_g\left(r\right)$ is given by the graviton  
geodesic equation~(\ref{nullgeo}).  

\begin{figure} 
\begin{center} 
\includegraphics[width=0.7\textwidth]{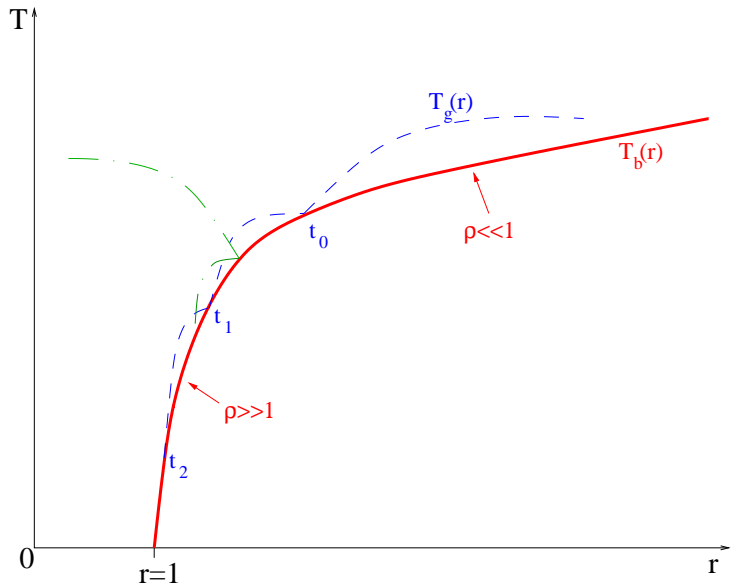} \\ 
\caption[gn_pic]{Trajectory of the brane and typical trajectories of 
gravitons in the
metric (14). The brane trajectory is the solid thick line, the
steepest part on the left corresponding to the high energy regime $\rho\gg
1$. The dashed line describes a graviton produced at $t_2$, bouncing
off the brane at $t_1$ and again at $t_0$. The dashed-dotted line
corresponds to a second graviton, that, after being reflected once by the
brane, falls into the bulk. } 
\end{center} 
\end{figure} 

Substituting  
\beq 
{}^{(1)}f_{(in)}=\frac{A}{2^{10}\,\pi^5}\,\kappa^2\, m_1^3 e^{-E_1/T_1}, 
\eeq 
in the integral (\ref{Tnn_in}), one finds, after integration over $p$, 
\beq 
{}^{(1)}{\cal {T}}_{nn}^{(in)}(\beta_0)= 
\frac{7!\,A}{2^{9}\,\pi^4}\,\kappa^2 
\int dx {\sinh^2x\sinh^3(\beta_1(x)-\beta_0+x)\over  
\cosh^8(\beta_1(x)-\beta_0+x)} \left({a_0\over a_1}\right)^{-5} T_1^8. 
\eeq 
 
In a similar way, one can compute the fraction ${}^{(2)}f_{in}$ due to the  
gravitons for which this is the second bounce on the brane since their  
emission.  
Consider such a graviton, which was emitted at time $t_2$, was reflected  
by the brane once at time $t_1$ and comes back again onto the brane  
at time $t_0$ ($t_2<t_1<t_0$).  
Just before the first bounce, the extra-dimensional momentum is given by 
\beq 
m_1^{(in)}\equiv p_1 \sinh x_1=- m_1^{(out)} 
=-{a_0\over a_1}p\sinh\left(\beta_1-\beta_0+x\right). 
\eeq 
Hence, at emission, we had  
\beq 
m_2^{(out)} 
={a_0\over a_2}p\sinh\left(\beta_2-2\beta_1+\beta_0-x\right). 
\eeq 
and thus  
\beq 
E_2^{(out)} 
={a_0\over a_2}p\cosh\left(\beta_2-2\beta_1+\beta_0-x\right). 
\eeq 
Consequently, the contribution to the transverse pressure is of the form 
\begin{eqnarray}\label{2reb} 
{}^{(2)}{\cal {T}}_{nn}^{(in)}(\beta_0)=&& 
\frac{7!\,A}{2^{9}\,\pi^4}\,\kappa^2 
\int dx {\sinh^2x\sinh^3(\beta_2(x)-2\beta_1(x)+\beta_0-x)\over  
\cosh^8(\beta_2(x)-2\beta_1(x)+\beta_0-x)}\cdot\nonumber\\ 
&&\cdot\left({a_0\over a_2(x)}\right)^{-5} T_2^8(x), 
\end{eqnarray}
where we have stressed the dependence on $x$ in the integrand (there is also  
a dependence on $\beta_0$).  
It is then straightforward to generalize to the contribution of gravitons  
with any number of bounces.

\section{Numerical results} 
 
\subsection{Source terms} 
 
Whereas the contributions from the newborn gravitons to the energy flux  
and pressure can be computed analytically, one must resort to  
numerical integration to compute the contribution from old gravitons.  
For the latter case, we need only the contribution to the pressure, since  
the contribution to the energy flux vanishes as explained earlier.  
Therefore, one can write the evolution equation for the dark component  
as 
\beq 
\dot\rho_D+4H\rho_D=\P^{(em)}-\P^{(b)}, 
\eeq 
with 
\beq\label{defp} 
\P^{(em)}=-2\left(1+\rho\right)\T^{(em)}_{un}-2H\T^{(em)}_{nn}, 
\qquad  
\P^{(b)}=4H\T^{(in)}_{nn}. 
\eeq 
The above equation can be rewritten as 
\beq\label{eqrhod} 
{d\over da}\left(a^4\rho_D\right)={a^3\over H}\P^{(em)}- 
{a^3\over H}\P^{(b)}\,\,, 
\eeq 
where $\P^{(em)}$ can be computed analytically by means of 
(\ref{Tnu}-\ref{Tnn}). 

We have computed numerically the function  
 $\P^{(b)}(a)$  using the generalization of Eq.~(\ref{2reb}), 
taking into account gravitons that have made up to $10^3$  
bounces before hitting the brane when its scale factor was $a$. We have  
checked that neglecting the effect of gravitons that have made more than  
$10^3$ bounces does not change appreciably our results. 
We have also assumed that 
the bulk contains no gravitons at the initial time $t_i$.
 
Numerically reliable results could be  
obtained for values up to  $\rho_i\simeq 10^3$. Taking 
 the lowest value  of $\mu$ compatible with 
the data (from small-scale gravity experiments), 
$\mu^{-1}\sim 0.1$ mm,  one gets that the 
brane tension has  
to be at least of the order of the TeV,
which  corresponds  to a fundamental  
Planck scale $M_5\equiv\kappa^{-2/3}\simeq 10^8$~GeV. 
As a consequence, the highest initial energy density which makes sense is
$\rho_i\sim M_5^4/\sigma\sim 10^{20}$. 

\begin{figure} 
\begin{center} 
\includegraphics[width=0.5\textwidth,angle=-90]{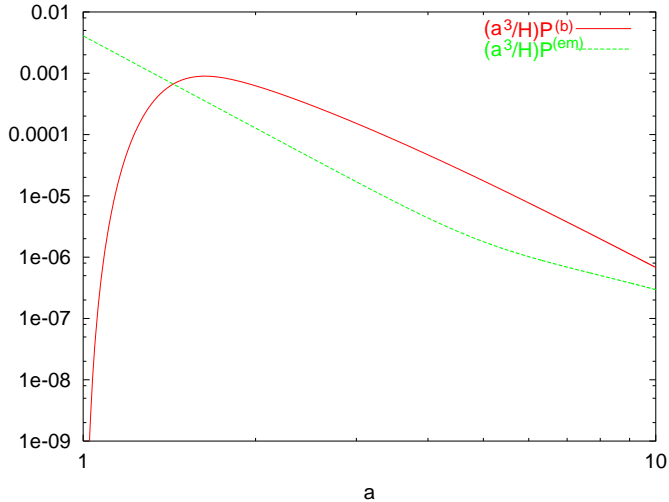} \\ 
\caption[gn_pic]{Contributions to the emission of 
dark radiation (for $\rho_i=1000$) from 
 graviton emission and from  the pressure
of bulk gravitons. ${\cal {P}}^{(b)}$ and ${\cal {P}}^{(em)}$ are defined in eq.~(\ref{defp}).} 
\end{center} 
\end{figure} 

In order to compare, at each instant in the history of the brane, the  
contribution directly due to the emission of gravitons with the one due  
to the reflection of older gravitons, we have plotted in  
 Fig. 2 the two terms on the right hand side of eq.~(\ref{eqrhod}). 
One observes that at early times the dominant contribution is that
due to the emission of  
gravitons, which is not surprising since the bulk is assumed to be empty  
initially. The bulk is then gradually filled with gravitons and some  
of them can come back onto the brane and  be reflected.  
They contribute to the transverse pressure effect, which can be seen  
on the plot to build very quickly. In the intermediate phase, the pressure  
effect dominates the emission effect so that the net source term for the  
dark radiation is negative. However, the two {\it integrated} effects 
are very  
close in amplitude, which means that the compensation observed in the  
simple Vaydia model is still working in this case.  
This is also the origin of the numerical difficulties of the present  
analysis: the result we are looking for is the small difference of 
two very large numbers. One can indeed see that both terms 
on the right hand side of eq.~(\ref{eqrhod}) scale 
roughly as $\rho_i^2$, whereas their difference
scales approximately as $\rho_i$. Notice that, for this reason, 
the quantities on the vertical axes of figures
2 and 3 have been rescaled by a factor $\rho_i^2$.

It is also important to stress the necessity to  take into account the 
multiple reflections of gravitons on the brane to get a correct evaluation 
of the total effect.
 To illustrate this point, we have plotted in Fig. 3
the contribution from gravitons for  
which this is the first bounce in comparison with the cumulative contribution  
from gravitons that have made up to $10^3$ bounces.

\begin{figure} 
\begin{center} 
\includegraphics[width=0.5\textwidth,angle=-90]{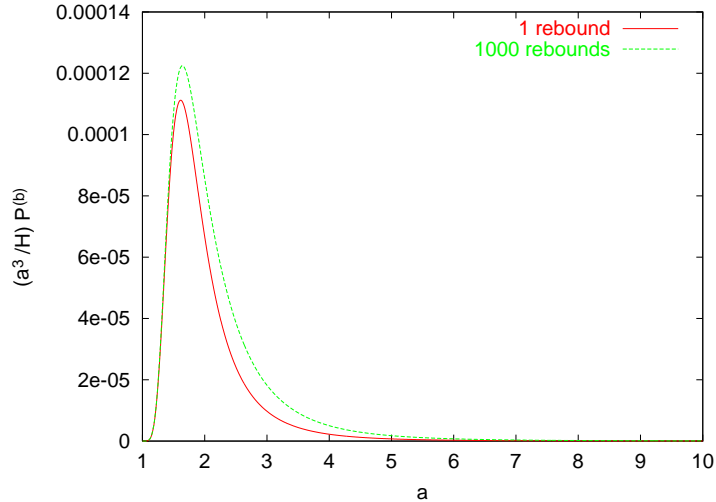} \\ 
\caption[gn_pic]{Contribution to the source term $(a^3/H)\,{\cal {P}}^{(b)}$ from 
gravitons that have been reflected only once by the brane, with respect 
to the total term (numerically, up to  $10^3$ bounces). } 
\end{center} 
\end{figure} 

\subsection{Dark radiation} 
Our main goal is to compute the dark radiation globally produced in the  
process. At very low energy, dark radiation is produced at  
a negligible rate, so that one can consider that there is an asymptotic  
constant value for the Weyl parameter $\C$. In the end, this asymptotic  
value for $\C$ depends only on the initial energy density in the brane and  
on the number of relativistic degrees of freedom during the high energy  
phase.
It will be convenient to express the final amount of dark radiation as  
its ratio  with respect to standard radiation energy density
\beq 
\epsilon_D\equiv {\rho_D\over \rho_{rad}}. 
\eeq 

Such quantity is constrained by cosmological observations. The amount of
nonstandard radiation in the early universe is usually measured in units of
extra neutrino species $\Delta N_\nu$. The conversion factor between $\Delta
N_\nu$ and $\epsilon_D$ is $\Delta N_\nu=\left(43/7\right)\,\left(g_*^{\mathrm
{nucl}}/g_*\right)^{1/3}\, \epsilon_D$, where $g_*^{\mathrm nucl}=10.75$ is the number of
relativistic degrees of freedom just  before  nucleosynthesis, more precisely
just before the electron-positon  annihilation. The factor $\left(g_*^{\mathrm
{nuc}}/g_*\right)^{1/3}$ accounts  for the change in the number of relativistic
degrees of freedom since the era in which dark radiation has been produced.
Assuming $g_*=106.75$,  this gives $\epsilon_D\simeq 0.35\,\Delta N_\nu$.
Constraints on $\Delta N_\nu$ depend upon the kind of observation one takes
into account, but a typical order of magnitude is $\Delta
N_\nu<0.2$~\cite{iykom02,bklms03}, which gives an upper limit
$\epsilon_D<0.07$. According  to both the analyses~\cite{iykom02,bklms03}, the
favored value of $\Delta N_\nu$ turns  out to be {\it negative}. Although this 
possibility is allowed by a negative value of the Weyl parameter $\C$, this
would correspond for the metric~(\ref{sch-ads}) to a naked singularity in the
bulk, a situation that is certainly not appealing from the theoretical point of
view. 

In the evaluation of the total amount of dark radiation produced, 
an important quantity is the energy loss due to production of gravitons, 
which depends on the number of relativistic degrees of freedom confined 
on the brane. 
In LSR,  the expression for the energy loss is given by
\beq\label{defalfa}
\sigma_B\equiv{\alpha\over 12}\kappa^2\rho^2, 
\eeq
where
\beq\label{alphabe}
 \alpha=
{ 212625\over 64\pi^7}\zeta(9/2)\zeta(7/2){\hat g\over g_*^2}
\eeq
with 
\beq
\hat g=\left((2/3)g_s+4g_v +\left(1-2^{-7/2}\right)\left(1-2^{-5/2}\right)
g_f\right), \qquad g_*=g_s+g_v+(7/8)g_f
\eeq
This is in exact agreement with the revised version~\cite{revhm} of HM.

In the present work , $\sigma_B=-{\cal {T}}^{(em)}_{un}$ and the expression 
for ${\cal {T}}^{(em)}_{un}$ is given in (\ref{Tnu}) so that 
$\alpha$, defined as in (\ref{defalfa}), is given by 
\beq
 \alpha=
{ 212625\over 64\pi^7}{A\over g_*^2}.
\label{alpha}
\eeq
The difference between the two above values for $\alpha$ comes from the fact
 that Fermi-Dirac or Bose-Einstein 
distribution functions were used in HM and LSR, whereas
here we have treated all the particles on the same footing and assume simply
a Boltzmann distribution. 
To make a precise comparison between the previous estimates and our numerical
results, we will use the analytical estimates of HM and LSR 
with the value for $\alpha$ given in (\ref{alpha}). For the particle content of the standard model, the values of $\alpha$ as given by eq.~(\ref{alphabe}) and~(\ref{alpha}) differ only by about the $5\%
$.

Let us now recall these previous estimates. 
In HM, the amount of dark radiation produced in the 
high energy regime can be expressed as
\beq
\epsilon_D\simeq {\cal F}{\alpha\over 4}\ln(\rho_i/2),
\eeq
where ${\cal F}$ is an  efficiency factor 
(denoted $\alpha$ in HM) with $5\pi/32<{\cal F}< 1$.
The amount of dark radiation produced during the whole low energy regime 
(assumed in HM to start at $\rho=2$) is $\alpha/2$. 
Therefore in our comparison we will take 
for  the total fraction of dark radiation as estimated in HM 
the expression
\begin{equation}\label{esthm}
\epsilon_D^{HM}=\frac{\alpha}{4}\,\left[2+{\cal F} \ln\left(\rho_i/2\right)\right]\,\,,\quad\rho_i\gg 2\,\,.
\end{equation}

The value of $\epsilon_D$ as given by LSR 
is obtained by solving a system of differential equations 
including~(\ref{eqlsr}). In the limit 
of large $\rho_i$ one gets $\epsilon_D\rightarrow \alpha/4$.

Let us now evaluate the amount of dark radiation produced in our 
numerical approach.
Starting from the equation 
\beq 
\dot\rho_D+4H\rho_D=\P^{(em)}-\P^{(b)}, 
\eeq 
where the source term is given explicitly above, one finds 
\beq 
\epsilon_D(t_0)={\rho_D(t_0)\over \rho_{rad}(t_0)}=  
{1\over \rho_{i}}\int_{t_i}^{t_0}dt \left({a(t)\over a_i}\right)^4 
\left(\P^{(em)}-\P^{(b)}\right). 
\eeq 
We have integrated numerically  this equation 
and plotted in Fig. 4 the evolution of the ratio  
 $\epsilon_D=\rho_D/\rho$ as a function of the scale
factor. Remarkably, for
large enough initial values of the energy density on the brane  $\rho_i$, 
 the dark radiation component can dominate the brane matter energy
density  at early times. 
This however does not seem to invalidate the implicit assumption that one
can neglect  the backreaction of bulk gravitons on the bulk geometry.
 Indeed, the dark energy domination, 
 $\rho_D\left(a\right)>\rho\left(a\right)$, is effective only
 in the high energy regime $\rho\gg 1$. And in this regime, the 
cosmological expansion is dominated by $\rho^2$, which remains much larger 
than $\rho_D$.
Nevertheless, it might be interesting  to explore the
consequences of the fact that in the very early evolution of the
Randall--Sundrum universe, most of the energy density of the universe should be
in the form of dark radiation (or bulk gravitons) gravitationally bound to the
brane. 

\begin{figure} 
\begin{center} 
\includegraphics[width=0.5\textwidth,angle=-90]{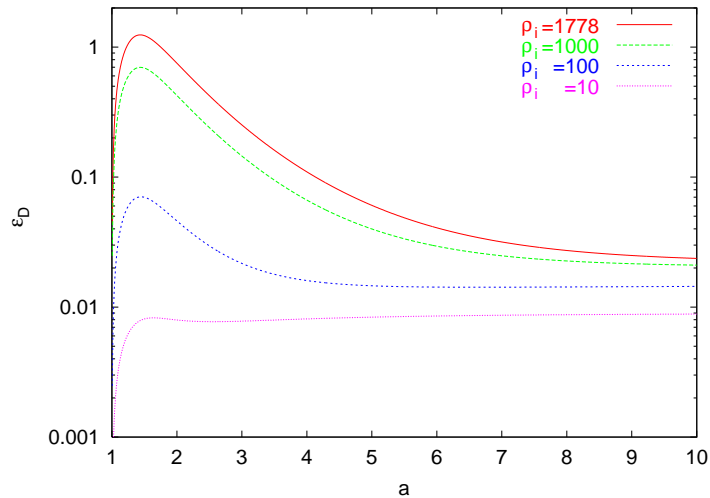} \\ 
\caption[gn_pic]{ Evolution of $\epsilon_D=\rho_D/\rho$ 
for different values of $\rho_i$.} 
\end{center} 
\end{figure} 

In Fig. 5, 
we  compare the numerical estimates of the present work with the analytical estimates of HM and LSR. We took for $\alpha$ the expression (\ref{alpha})
with all the degrees of freedom of the standard model.
 The curve corresponding to HM comes 
from eq.~(\ref{esthm}) with the lowest value of ${\cal {F}}$.

Due to the difficulties mentioned above,
 the range of our numerical analysis is limited to values 
of $\rho_i$ smaller than about $2\times 10^3$. 
In this range, 
we get an accurate description of the behavior 
of $\epsilon_D$ as a function of $\rho_i$. 
 We see that $\epsilon_D$ is a slowly increasing function of $\rho_i$.
At low values of $\rho_i$, the numerical results are close to
 the estimate of LSR . 
Indeed, for $\rho_i \la 1$, the
effect of graviton bounces can be neglected, and the Vaidya description
analyzed in section 5 gives a good approximation, whose results agree, in
the limit $\rho_i\ll 1$, also with those of HM.
Going to higher values of $\rho_i$, in the regime of validity of our analysis the value of $\epsilon_D$ 
is below the lowest bound estimated by HM, but it gets closer and closer 
as $\rho_i$ increases and 
 one can expect it to become higher 
already for $\rho_i$ of the order of few thousands 
(remember that the maximal value of $\rho_i$ compatible 
with constraints on the Randall-Sundrum model is 
$\rho_i\sim 10^{20}$). It is remarkable that already 
for the (relatively) low value $\rho_i\simeq 1800$, 
$\epsilon_D>0.02$, which  is not far from  the upper bound imposed 
by CMB and BBN observations~\cite{iykom02,bklms03}.

\begin{figure} 
\begin{center} 
\includegraphics[width=0.5\textwidth,angle=-90]{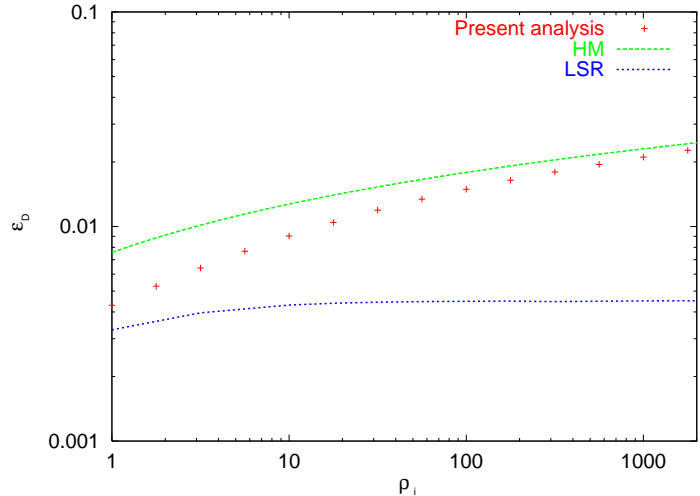} \\ 
\caption[gn_pic]{Comparison of the numerical results of the present work with the estimates of HM (upper curve) and of LSR (lower curve) for the amount of dark radiation $\epsilon_D=\rho_D/\rho$ as a function of the initial energy density on the brane $\rho_i$.}
\end{center} 
\end{figure} 
 
\section{Conclusions}
In the present work, we have computed  the amount of 
dark radiation produced 
during the cosmological  evolution of our brane-universe.
Previous estimates were based on relatively crude approximations: in one
case, the evolution of the brane was separated into a high energy phase 
and a low energy phase; in the other case, all bulk gravitons were 
supposed to be radial. However, both estimates were pointing to
 values very close to the current bound on extra relativistic degrees
of freedom, which was motivating a more detailed study of the question.

To go beyond the previous approximations, the present work uses 
a  numerical approach, which enables us to treat smoothly the transition 
between the high and low energy regimes and to deal with non radial 
gravitons. Our initial  objective to compute precisely the amount 
of dark radiation produced is however hampered by a problem of numerical 
precision, and we have been able to do this computation only for moderate
values of the initial brane energy density. The reason for this limitation
lies in a remarkable compensation between two opposite effects: the 
emission of bulk gravitons by the brane, which contributes {\it positively} to 
the dark radiation, and the pressure of old gravitons bouncing off the 
brane, which contributes {\it negatively} to the dark radiation. 
Numerically, we have been obliged to compute these two effects separately,
and the net effect, in which we are interested, comes from the difference 
of two quasi-equal huge numbers, which is difficult to control numerically.

With the present analysis, we could estimate the evolution of 
the dark radiation component as a function of time, finding 
that in the very early stages of a radiation dominated 
brane universe, energy as dark radiation could 
easily exceed the amount of energy in brane matter. 

In the range of initial densities where we can rely on our numerical 
computation, we have been able to compare our result with the analytical 
estimates obtained previously and our results 
agree with their order of magnitude. 

The increase  of the amount 
of dark radiation produced with  the initial energy density 
on the brane seems to indicate that  a significant amount of dark 
radiation will be produced for  an extended high energy (non standard)
 period in the early universe. However, we cannot safely
 extrapolate the present analysis 
to higher values of the initial energy density on the brane and we believe 
that 
an extension of our  numerical computation further into
 the very high energy regime will tell us whether 
a long period of nonstandard cosmology is going to be soon 
ruled out  by observations.

{\bf Acknowledgements}
We wish to thank T. Tanaka for very stimulating discussions at an early 
stage of the present work. The work of L.S. is supported by the European 
Community's Human Potential Programme under contract HPRN-CT-2000-00152
Supersymmetry and the Early Universe.

\end{document}